\documentclass[manuscript]{aastex}
\usepackage{emulateapj5}

\def\HST{{\it HST}}
\def\kms{{\>\rm km\>s^{-1}}}

\def\Spitzer{{\it Spitzer}}

\shorttitle{Luminous Transient in NGC 300}
\shortauthors{Bond et al.}

\begin{document}

\title{The 2008 Luminous Optical Transient in the Nearby Galaxy
NGC~300\altaffilmark{1}}

\author{
Howard E. Bond,\altaffilmark{2} 
Luigi R. Bedin,\altaffilmark{2}
Alceste Z. Bonanos,\altaffilmark{2,3}
Roberta M. Humphreys,\altaffilmark{4}
L.A.G. Berto Monard,\altaffilmark{5}
Jos\'e L. Prieto,\altaffilmark{6}
and
Frederick M. Walter\altaffilmark{7}
}

\altaffiltext{1}  
{Based in part on observations with the NASA/ESA {\it Hubble Space Telescope\/}
obtained at the Space Telescope Science Institute, and from the data archive at
STScI, which is operated by the Association of Universities for Research in
Astronomy, Inc., under NASA contract NAS5-26555; in part on archival data
obtained with the {\it Spitzer Space Telescope}, which is operated by the Jet
Propulsion Laboratory, California Institute of Technology, under a contract with
NASA; in part on observations obtained with the 6.5-m Magellan Clay Telescope
located at Las Campanas Observatory, Chile; and in part on observations obtained
with the SMARTS Consortium 1.3- and 1.5-m telescopes located at Cerro Tololo
Interamerican Observatory, Chile.}

\altaffiltext{2}
{Space Telescope Science Institute, 3700 San Martin Dr., Baltimore, MD
21218; [bond, bedin, abonanos]@stsci.edu}

\altaffiltext{3}
{Giacconi Fellow}

\altaffiltext{4}
{Astronomy Department, University of Minnesota, Minneapolis, MN 55455;
roberta@aps.umn.edu}

\altaffiltext{5}
{Bronberg Observatory, PO Box 11426, Tiegerpoort 0056, South Africa}

\altaffiltext{6}
{Department of Astronomy, The Ohio State University, 140 W. 18th Ave., Columbus,
OH 43210; prieto@astronomy.ohio-state.edu}

\altaffiltext{7}
{Department of Physics \& Astronomy, Stony Brook University, Stony Brook, NY
11704; fwalter@astro.sunysb.edu}

% \clearpage

\begin{abstract}

A luminous optical transient (OT) that appeared in NGC~300 in early 2008 had a
maximum brightness, $M_V\simeq-12$ to $-13$, intermediate between classical
novae and supernovae. We present ground-based photometric and spectroscopic
monitoring and adaptive-optics imaging of the OT, as well as pre- and
post-outburst space-based imaging with \HST\/ and {\it Spitzer}. The optical
spectrum at maximum showed an F-type supergiant photosphere with superposed
emission lines of hydrogen, \ion{Ca}{2}, and [\ion{Ca}{2}], similar to the
spectra of low-luminosity Type~IIn ``supernova impostors'' like SN~2008S, as
well as cool hypergiants like IRC~+10420. The emission lines have a complex,
double structure, indicating a bipolar outflow with velocities of
$\sim$$75\,\kms$. The luminous energy released in the eruption was
$\sim$$10^{47}$ ergs, most of it emitted in the first 2 months. By registering
new \HST\/ images with deep archival frames, we have precisely located the OT
site, and find no detectable optical progenitor brighter than broad-band $V$
magnitude 28.5.  However, archival {\it Spitzer\/} images reveal a bright,
non-variable mid-IR pre-outburst source. We conclude that the NGC~300~OT was a
heavily dust-enshrouded luminous star, of  $\sim$10--$15\,M_\odot$, which
experienced an eruption that cleared the surrounding dust and initiated a
bipolar wind. The progenitor was likely an OH/IR source which had begun to
evolve on a blue loop toward higher temperatures, but the precise cause of the
outburst remains uncertain.

\end{abstract}


% \clearpage

% {\bf New ApJL limits: 3500 words, abstract 250 words, 5 total figs.\ \&
% tables.}

\section{Introduction: Transients in the Nova--Supernova Gap}

In recent years, eruptive objects with maximum luminosities intermediate between
those of classical novae (CNe) and supernovae (SNe) have been discovered in
increasing numbers, as surveys for Galactic and extragalactic transients are
made with greater depth and sky coverage by professional and amateur
astronomers. The term ``supernova impostors'' was introduced by Van Dyk et al.\
(2000) following the outburst of SN~1997bs. SN~1997bs was classified as a SN~IIn
event, ``n'' denoting narrow emission lines during outburst, in contrast to the
high ejection velocities of typical SNe. However, it reached an absolute
magnitude at maximum of only $M_V\simeq-13.8$, more than 3~mag fainter than a
typical core-collapse SN\null. Van~Dyk et al.\ argued that SN~1997bs was instead
a ``superoutburst'' of a massive luminous blue variable (LBV), analogous to
those experienced by the Galactic objects P~Cygni and $\eta$~Carinae (see
Humphreys, Davidson, \& Smith 1999). The subject of intrinsically faint SNe~IIn
and $\eta$~Car analogs has been summarized by Van~Dyk (2005) and Smith et al.\
(2008). More recently, SN~2008S, which had a SN~IIn-type spectrum but reached
only $M_R\simeq-13.9$, has been discussed by Thompson et al.\ (2008) and Smith
et al.\ (2008). Prieto et al.\ (2008) have identified the precursor of SN~2008S
in pre-outburst mid-IR images obtained with the {\it Spitzer Space Telescope},
demonstrating that the progenitor was a luminous star deeply embedded in
circumstellar dust.

V838~Monocerotis is the prototype of an apparently separate class of transients
also lying in the CN-SN gap. V838~Mon erupted in 2002, illuminating a
spectacular light echo that provided a geometric distance (Sparks et al.\ 2008)
implying an absolute magnitude at maximum of $M_V=-9.8$. Unlike a CN, V838~Mon
became progressively redder during its outburst, eventually becoming enshrouded
in circumstellar dust. Its outburst light curve (Bond et al.\ 2003) was very
different from that of a SN or LBV eruption, showing a series of four maxima
separated by about a month each. The cause of this outburst remains uncertain,
but a leading suggestion is that it was a stellar collision or merger (e.g.,
Tylenda \& Soker 2006, Corradi \& Munari 2007, and references therein). 

Possible extragalactic analogs of V838~Mon---although this interpretation is
controversial---include the ``M31~red variable'' (``M31~RV'') which reached
$M_{V\rm,max}\simeq-9.4$ (Bond \& Siegel 2006 and references therein), and a
luminous red transient in M85 (Rau et al.\ 2007; Kulkarni et al.\ 2007), which
attained $M_{V\rm,max}\simeq-13$. M31~RV and the M85 transient appear to have
occurred in old populations, suggesting that they did not arise from very
massive progenitors.

In this {\it Letter}, we report observations of a luminous transient discovered
in 2008 May in the nearby spiral galaxy NGC~300. We obtained extensive 
ground-based imaging, photometry, and spectroscopy, and we also investigated
pre- and post-outburst space-based optical and infrared images of the
transient.  This {\it Letter} presents an overview of our results, along with a
preliminary discussion of the nature of this remarkable object. Subsequent
publications will give fuller details. We designate the object as the ``NGC~300
optical transient,'' hereafter ``NGC~300~OT.'' 

\section{Discovery and Outburst Light Curve}

The outburst of NGC~300~OT was discovered by Monard (2008) during his SN search
program with the Bronberg Observatory 0.3-m telescope equipped with a CCD
camera. The OT was detected at broad-band magnitude 14.3 on a frame obtained on
2008 May 14 (all dates in this paper are UT)\null. Subsequent inspection of an
image from 2008 April 24, taken in morning twilight shortly after NGC~300 had
emerged from behind the Sun, showed that the OT was already rising, at
$\sim$16.3~mag. It was fainter than 18~mag on 2008 February~8, and on all
previous Bronberg observations.

NGC~300 is an SAd spiral in the Sculptor Group, located just outside the Local
Group. The OT lies in a spiral arm with active star formation. We adopt a
distance modulus of $(m-M)_0=26.37$ ($d=1.88$~Mpc), based on Cepheids (Gieren et
al.\ 2005) and the red-giant tip (Rizzi et~al.\ 2006).

At maximum the OT was much fainter than a typical SN, but brighter than any
CN\null. Because of our interest in luminous red transients, we began a program
of photometric and spectroscopic monitoring, using the 1.3- and 1.5-m telescopes
at Cerro Tololo Interamerican Observatory operated by the SMARTS
Consortium.\footnote{SMARTS is the Small and Medium Aperture Research Telescope
System; see http://www.astro.yale.edu/smarts/} The photometric monitoring in
particular was intensive at first, because of our expectation that the OT might
exhibit periodic spikes, like those during the outburst of V838~Mon (Bond et
al.\ 2003).

We used the ANDICAM optical/near-IR direct camera (DePoy et al.\ 2003) on the
SMARTS 1.3-m with CCD and IR detectors, operated simultaneously, for the {\it
BVRI\/} and {\it JHK\/} photometry. The {\it BVRI\/} magnitudes of the OT were
determined differentially with respect to a nearby comparison star,  calibrated
on photometric nights using Landolt (1992) standard fields. The {\it JHK\/}
magnitudes are differential with respect to the same star, calibrated using its
2MASS photometry. The J2000 position of the OT, based on astrometry of ANDICAM
frames calibrated against the USNO-NOMAD catalog, is RA = 00:54:34.51, Dec =
$-37$:38:31.4, with errors in each coordinate of about $\pm$$0\farcs2$.

Figure~1 shows the {\it BVRIJHK\/} light curve. Unfortunately, the rise to
maximum was poorly covered, but it appears that it was more rapid than the
subsequent slower decline. The brightest $V$ magnitude we measured was 14.69, on
the first night of SMARTS observations (2008 May 15), corresponding to
$M_{V,0}=-12.0$ to $-12.9$ for the adopted distance and a reddening lying in the
range $E(B-V)=0.1$ to 0.4 (see \S3.1). Following maximum light, the OT has
declined smoothly in brightness at all wavelengths, while becoming steadily
redder. For example, $V-K$ evolved from 3.1 in mid-May to 7.4 at the beginning
of 2008 December. The rate of decline in the optical lessened around the
beginning of 2008 September, but at this writing the brightness continues to
decrease in both the optical and near-IR\null. The $R$ magnitudes are not fading
as rapidly as in the other filters, due to strong H$\alpha$ emission.

\section{Spectroscopy}

\subsection{Low Resolution}

We obtained low- and moderate-resolution optical spectroscopy of NGC~300~OT
throughout its outburst, using the SMARTS 1.5-m Cassegrain spectrograph.  Our
first low-resolution spectrum was observed on 2008 May 15 (Bond, Walter, \&
Vel\'asquez 2008) and is shown in Figure~2 (top spectrum; resolution
17.2~\AA)\null. This spectrum exhibits strong emission at H$\alpha$, H$\beta$,
the \ion{Ca}{2} triplet at 8542--8498--8662~\AA, and the unusual forbidden
[\ion{Ca}{2}] doublet at 7291--7323~\AA\null. The Balmer lines are only slightly
resolved at the velocity resolution ($790 \kms$) of the spectrum. The underlying
continuum resembles a high-luminosity F-type supergiant, with \ion{Ca}{2} H and
K seen in absorption along with several weaker luminosity-sensitive lines such
as the \ion{O}{1} triplet at 7774~\AA\null. Our photometry at maximum and
shortly afterwards yields $B-V\simeq0.8$, implying that the inter- and
circumstellar extinction could be as high as $E(B-V)\simeq0.4$, since a mid- to
late F-type supergiant normally has $(B-V)_0\simeq0.4$. The mean foreground
reddening of NGC~300 is $E(B-V)\simeq0.1$ (Gieren et al.\ 2005).

This spectrum is clearly not that of a CN nor a typical SN\null. With strong
\ion{Ca}{2} and [\ion{Ca}{2}] emission and an F-type absorption spectrum, it
resembles that of the Type~IIn supernova impostor SN~2008S (Steele et al.\ 2008;
Smith et al.\ 2008). It is also closely similar to the spectra of the
well-studied post-red supergiant IRC~+10420 (Jones et al.\ 1993; Humphreys,
Davidson, \& Smith 2002), Variable~A in M33 in its recent quiescent state
(Humphreys et al.\ 2006), and V838~Mon near its maximum (e.g., Wisniewski et
al.\ 2003). These F-type supergiant spectra with superposed narrow emission
lines of relatively low excitation are indicative of eruptions that produce an
optically thick wind or slowly expanding envelope. 

As the OT declined, its spectrum became dominated by the H$\alpha$ and
\ion{Ca}{2} emission. One example, selected from our monitoring program, is the
bottom spectrum shown in Figure~2, obtained on 2008 August~23. By this date, the
continuum had faded dramatically and reddened somewhat, but the strong emission
remained.

\subsection{Moderate Resolution}

We also obtained moderate-resolution (3.1~\AA) spectra regularly with the SMARTS
1.5-m spectrograph, and on three occasions with the Magellan 6.5-m telescope and
MagE echellette spectrograph (Marshall et al.\ 2007; resolution 1.5~\AA\ at
\ion{Ca}{2}, 1.1~\AA\ at H$\alpha$): 2008 July~6, Aug 30, and Sep~1. These
spectra show that the \ion{Ca}{2} triplet, H$\alpha$, and H$\beta$ have
double-peaked emission lines, with the blue component being the stronger. 

Figure~3 shows the H$\alpha$ and \ion{Ca}{2} emission features from the Magellan
echellette spectra on 2008 July~6 (top spectrum) and the average of the August
30 and September~1 observations (bottom spectrum), and illustrates the double
structure of these lines. Double emission is usually attributed either to a
bipolar outflow or a rotating disk, but in the case of an eruption, the double
lines are most likely formed in a bipolar wind.  IRC~+10420 shows very similar
double-peaked emission and has strong independent evidence for a bipolar outflow
(Davies et al.\ 2007; Patel et al.\ 2008). 

The peak of the OT's primary blue component slowly increases in strength with
time relative to the red component, supporting our conclusion that the outflow
is bipolar, with the nearer, blue-shifted lobe expanding towards us. The Doppler
velocities measured from both the hydrogen and \ion{Ca}{2} emission lines in the
echellette spectra indicate expansion velocities for the primary wind of 
$\sim$70--$80 \kms$ with respect to the star's systemic velocity of about
$+196\kms$. There is evidence as well for additional emission components,
appearing as shoulders or secondary bumps on the primary blue and red peaks at
velocities of $\approx$$160\kms$. 

% The systemic velocity, measured from the absorption lines, is compatible with
% the OT's expected velocity of $+190 \kms$, based on NGC 300's rotation curve 
% at its location in the galaxy (Marcelin, Boulesteix, \& Georgelin 1985). 

A wind velocity of $\sim$$75 \kms$ is typical of the winds of F-type
supergiants, and is somewhat lower than the winds associated with LBVs in
eruption. It is very similar to the expansion velocity of $\sim$$60\kms$ of
IRC~+10420 (Jones et al.\ 1993; Humphreys et al.\ 2002). By contrast, the
[\ion{Ca}{2}] lines, formed in a very low-density region, do not show a
double-peaked structure; they probably arise in more-distant and slower-moving
material, likely formed before the current eruption. 

\section{\HST\/ and VLT Pre- and Post-Outburst Imaging}

We imaged the site of the OT twice with the {\it Hubble Space Telescope\/}
(\HST), in the Director's Discretionary (DD) program GO-11553 (PI: Bond). We
used the Wide Field Planetary Camera~2 (WFPC2) with the F450W and F814W filters;
observations were made on 2008 June~9 and September~1. In addition, the \HST\/
archive contains two sets of deep observations made before the outburst,
obtained with the Wide Field Channel of the Advanced Camera for Surveys (ACS)
and the F435W, F475W, F555W, F606W, and F814W filters. The ACS observations were
made on 2002 December~25 (GO-9492, PI: F.~Bresolin) and 2006 November 8--10
(GO-10915, PI: J.~Dalcanton). We used these data to locate the OT precisely,
search for a progenitor object, study the surrounding stellar population, and
set limits on a light echo from the outburst.

In addition to the \HST\/ data, we were awarded DD time on the ESO Very Large
Telescope (VLT) to obtain near-IR imaging of the OT, using the adaptive-optics
NaCo camera (Lenzen et al.\ 2003; Rousset et al.\ 2003). The VLT\slash NaCo
images were obtained on 2008 June~6. 

\subsection{Astrometry and Pre-Outburst Magnitude Limits}

The WFPC2 exposure times for our DD observations ranged from 5 to 260~s, chosen
so as not to saturate the image of the OT, but to show enough of the surrounding
star field for precise registration with the pre-outburst ACS frames. The
archival ACS frames are considerably deeper, with total exposure times of 1080,
2976, 1080, 3030, and 4524~s in F435W, F475W, F555W, F606W, and F814W,
respectively.  The VLT/NaCo $J$-band frames were four 600~s exposures, combined
in the ESO pipeline to produce a single image.

Figure~4 shows a montage of post- and pre-outburst images.  The~OT lies amid a
rich field of resolved stars. Remarkably, however, there is no significant
detection of a progenitor object at the OT's precise location. The two nearest
reliably detected sources lie about $0\farcs15$ and $0\farcs25$ from the site,
well in excess of our registration error of about $0\farcs01$. These two stars
are red giants with F606W magnitudes of about 26.3 and 27.0. In the deepest
stacked pre-outburst images there are suggestions of local maxima at the OT
site, but these are all consistent with noise. From the three deepest ACS
stacked images, we find the following limiting magnitudes (Vega-mag scale,
3$\sigma$ upper limits) for the OT progenitor: 28.3 (F475W), 28.5 (F606W), and
26.6 (F814W).  Similar limiting magnitudes were reported earlier, based on
registering the 2006 ACS images with ground-based Magellan/Clay 6.5-m images, by
Berger \& Soderberg (2008), but with larger positional uncertainties. 

The OT was thus very unlike a typical core-collapse SN, for which luminous
progenitor stars are being found relatively easily in archival \HST\/ images at
distances greater than that of NGC~300 (e.g., Smartt et al.\ 2008 and references
therein), nor was it similar to erupting LBVs, which even in quiescence have
high optical luminosities (e.g., Humphreys \& Davidson 1994).

\subsection{The Surrounding Young Stellar Population}

The \HST\/ ACS images also provide deep CMDs for the environment surrounding
the~OT\null. These will be presented in detail in a subsequent publication, but
we note that there is a rich population of young objects in the immediate
vicinity of the OT\null.  For example, within $2\farcs5$ ($\sim$23~pc) of the OT
site there are over a dozen blue main-sequence stars brighter than about F606W
magnitude 26, reaching as bright as $M_V\approx-3$ (corresponding to B1 stars of
$\approx$$14\,M_\odot$)\null. The surrounding spiral-arm field is rich in blue
clusters and associations.

% 12/30/08: brightest MS stars in CMD (red circles) have m(814) = m(606) =
% 23.6. Mv is 23.6-26.37-3.1*0.096=-3.1.  Corresponding SpT is about B1;
% corresp mass is roughly 14 Msun.

\subsection{Search for Light Echo}

We compared the post-outburst \HST\/ images of 2008 June~9 and September~1 in
search of an expanding light echo, similar to the one surrounding V838~Mon
(e.g., Bond 2007, Sparks et al.\ 2008, and references therein). Although a light
echo of the surface brightness and angular size of V838~Mon's, moved to the
distance of NGC~300, would have been detectable at \HST\/ resolution, no echo
is seen.

\section{\Spitzer\/ Pre-Outburst Observations: An Enshrouded Red Supergiant}

The very faint limits on the pre-outburst optical brightness of NGC~300~OT
initially suggested to us that there was a similarity with V838~Mon, whose
progenitor was also optically inconspicuous (e.g., Af{\c s}ar \& Bond 2007, and
references therein). However, the very different light curve and spectroscopic
behavior of the OT as its outburst proceeded casts doubt on this parallel.

Moreover, a striking difference emerged when one of us (Prieto 2008) reported
that the progenitor of the OT {\it was\/} detected as a bright mid-IR source in
archival observations obtained with {\it Spitzer\/} on 2003 November~21 (PI:
G.~Helou) and 2007 December~28 (PI: R.~Kennicutt). The source was present in all
four IRAC bands (3.6, 4.5, 5.8, and 8~\micron), and in the MIPS 24~\micron\
band. A black-body fit to the 3.6--24~\micron\ fluxes gives the following
results for the luminosity, effective temperature, and radius of the OT
progenitor: $L_{\rm bb} = 5.5\times10^4\, L_\odot$, $T_{\rm bb} = 350$~K, and
$R_{\rm bb} = 300$~AU\null. These are consistent with a heavily dust-enshrouded
red supergiant of $\sim$10--$15\,M_\odot$ (e.g., Maeder \& Meynet 2000). There
was no variability of the progenitor between 2003 and 2007. Further detailed
discussion of the {\it Spitzer\/} observations is given by Thompson et al.\
(2008).

% the first observation of N300 with Spitzer IRAC imaging was obtained 11/21/2003, 
% the ID of the proposal is 1083 and the PI was G. Helou. The info you have in
% the paper is for the last one obtained in Dec. 2007. So there are 2 epochs 
% separated by 4 years for N300, and there are several epochs starting 
% 3 years before explosion for the progenitor of SN2008S (you can see
% the info in the paper).

\section{Nature of the Luminous Transient in NGC 300}

A picture thus emerges in which a luminous massive star, optically inconspicuous
because it was deeply embedded in a dusty envelope, underwent a sudden outburst
that cleared most of the surrounding dust.  In many respects, NGC~300~OT has
behaved remarkably similarly to the supernova impostor SN~2008S (see \S1).
SN~2008S was also heavily enshrouded by dust prior to its outburst, with an
intrinsic luminosity of $3.5 \times 10^4 \, L_\odot$ based on its IR SED,
implying an initial mass of $\sim$$10\,M_\odot$\null. 

The energy released in the OT's eruption was relatively modest. Integrating over
the extinction-corrected visual light curve, beginning with the initial
detection in 2008 April, and assuming that the bolometric correction is zero, we
find that the total energy released was $\sim$$0.8$--$2 \times 10^{47}$~ergs.
This is similar to the energy radiated by SN~2008S, $\sim$$6\times10^{47}$~ergs
(Smith et al.\ 2008), but considerably less than the $\sim$$10^{50}$ ergs
emitted in $\eta$~Car's great eruption, and of course the $\sim$$10^{51}$ ergs
in true core-collapse SNe.

Given the large amounts of pre-outburst obscuring dust, both NGC~300~OT and
SN~2008S must have experienced significant mass loss as post-main-sequence
stars, and were most likely red supergiants (RSGs), post-RSGs, or conceivably
lower-mass stars at the tip of the AGB just before the outbursts. Thompson et
al.\ (2008) have suggested that these events may represent a new class of
low-luminosity outbursts arising from stars in the mass range $\sim$8--$11\,
M_\odot$\null. By comparison with multi-epoch {\it Spitzer\/} IRAC observations
in M33, they have identified a group of rare, reddened sources that they call
extreme-AGB stars, which may be the progenitor class in a short-lived stage
lasting only $\le$$10^4$~yr immediately preceding the outbursts.  There are also
many enshrouded OH/IR stars in our own Galaxy, including AGB-tip stars and red
supergiants with very red mid-IR colors; some may be as red as the progenitors
of the OT and SN~2008S.

% Their luminosities imply relatively high masses, $\sim$$10\,M_{\odot}$, but both
% stars were at or close to the AGB--limit. and thus might have been lower-mass
% stars.

Here we suggest that NGC 300 OT, and by analogy SN~2008S, were OH/IR stars
before their eruptions. Given both objects' lack of obvious variability from
{\it Spitzer\/} observations several years apart, these OTs were not Mira
variables just before outburst.  They had presumably left the region of RSGs and
AGB stars in the HR diagram, and were on a blue loop back to warmer
temperatures. The NGC~300 OT's bipolar wind at $\approx$75 km s$^{-1}$ also
supports a warmer temperature for the progenitor.  During this transition, both
stars reached a state in which they exceeded the Eddington limit for their
luminosities and masses and suddenly initiated outflows, bipolar in the case of
NGC~300~OT\null. Exactly why this should happen remains uncertain: the eruptions
may have resulted from some type of as-yet unexplained failed supernova, a
binary merger, or a photospheric eruption.

% During their post-RSG evolution, they may encounter a temperature regime
% (6000--9000~K) of increased dynamical instability (de~Jager 1998), driven by
% increasing opacity possibly in combination with increasing rotation and
% pulsation. 

% The presence of \ion{Ca}{2} emission, observed in the winds of F-type
% hypergiants, suggests that the progenitor was not as hot as a B-type supergiant,
% as in the case of LBV eruptions. The implication that the underlying star in
% NGC~300~OT  was somewhat warmer than an RSG or AGB star is also supported by its
% bipolar wind with velocities of $\sim$$75\,\kms$. Such velocities are
% significantly higher than the winds associated with RSGs, AGB stars, and OH/IR
% stars. The [\ion{Ca}{2}] emission, however, originates in slower-moving,
% lower-density gas, possibly a remnant from the progenitor's earlier state, such
% as an RSG.

Continued observations of the NGC~300~OT as it fades are especially important.
We currently observe the spectrum of its wind, but if it does not form 
another cocoon of dust, we may eventually observe the survivor, determine its
true nature, and possibly gain some insight into the cause of its eruption. 

\acknowledgments

Partial support for this research was provided by NASA through grant  GO-11553
from the Space Telescope Science Institute, which is operated by the Association
of Universities for Research in Astronomy, Inc., under NASA contract NAS5-26555.
STScI's participation in the SMARTS Consortium is supported by the STScI
Director's Discretionary Research Fund. We thank M.~Hern\'andez  and
J.~Vel\'asquez for conducting the 1.5-m observations, J.~Espinoza,  A.~Miranda,
and J.~Vasquez for the 1.3-m observations, and Michelle Buxton and Suzanne
Tourtellotte for scheduling the 1.3-m telescope. G.~Pietrzy{\'n}ski kindly
obtained one of the MagE spectra. We thank Kris Davidson, Chris Kochanek, Kris
Stanek, and Todd Thompson for useful discussions. 

{\it Facilities:} \facility{HST (WFPC2, ACS)},  \facility{Magellan:Clay},
\facility{SMARTS:1.5m},  \facility{SMARTS:1.3m}, \facility{Spitzer},
\facility{VLT:NaCo}.

% \clearpage

\begin{figure}[hbt]
\begin{center}
\includegraphics[width=5in]{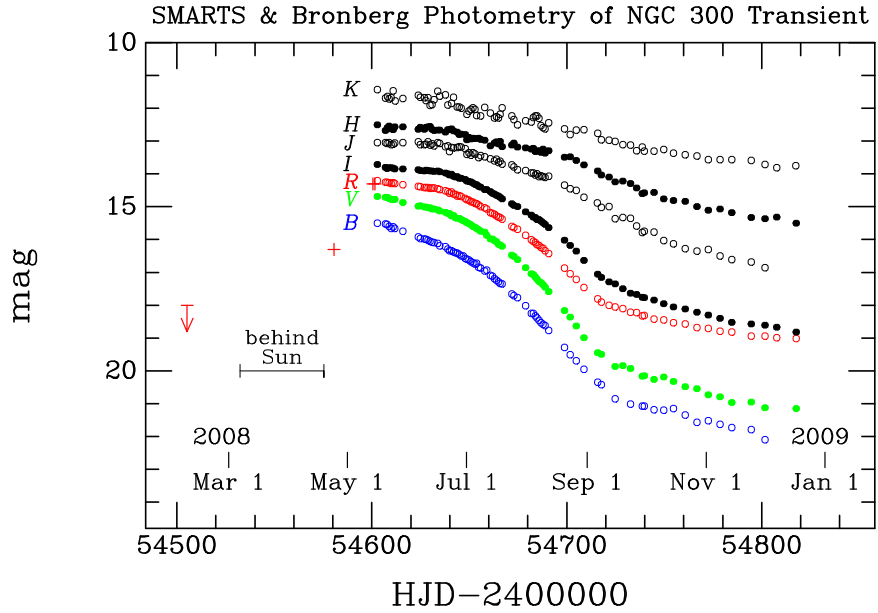}
\end{center}
\figcaption{
{\it BVRIJHK\/} light curve of the NGC~300 optical transient. SMARTS 1.3-m data
are shown as open and filled circles; Bronberg discovery observations and the
pre-discovery detection are shown as crosses (these broad-band magnitudes are
close to Landolt $R$), and the downward arrow on the left shows the upper limit
in 2008 February. 
}
\end{figure}

\begin{figure}
\begin{center}
\includegraphics[width=5in]{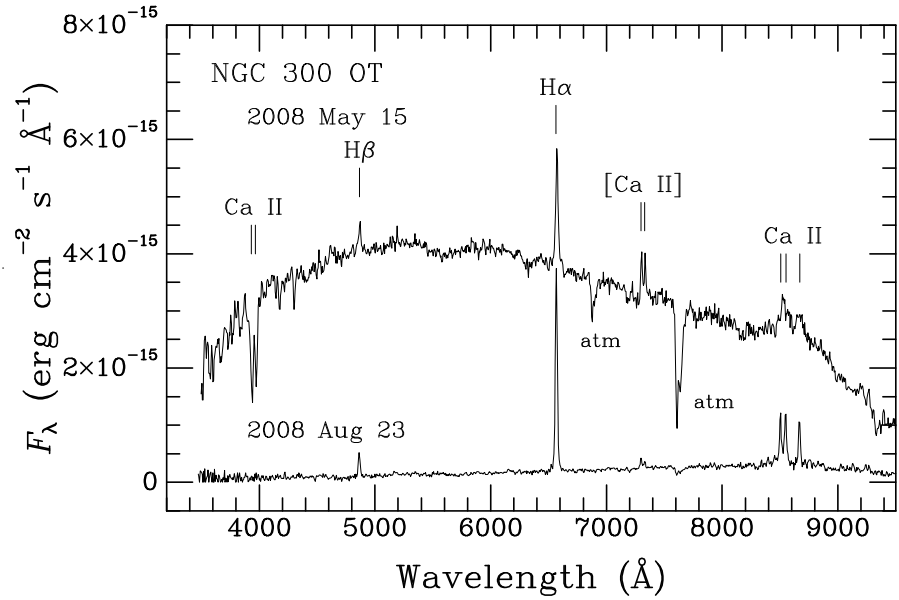}
\end{center}
\figcaption{
Low-resolution ($\sim$17.2~\AA) SMARTS 1.5-m spectra of the OT on 2008 May 15
and August 23. In the May~15 observation, note strong emission lines of H,
\ion{Ca}{2}, and [\ion{Ca}{2}], superposed on an F-type absorption spectrum. By
August, the continuum had faded dramatically and the spectrum is dominated by
the emission lines.
}
\end{figure}

\begin{figure}
\begin{center}
\includegraphics[width=5in]{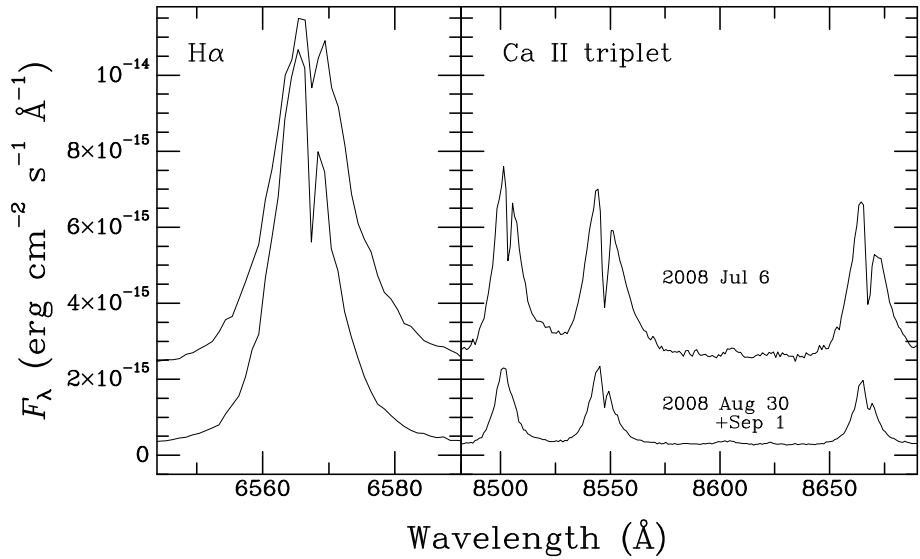}
\end{center}
\figcaption{
Portions of Magellan 6.5-m MagE echellette spectra (resolution $\sim$1.3~\AA) of
the OT on 2008 July~6 and the average of two spectra taken on August 30 and
September~1. The emission lines of H$\alpha$ and the \ion{Ca}{2} triplet are
double, indicating a bipolar outflow.
}
\end{figure}

\begin{figure}
\begin{center}
\includegraphics[width=\columnwidth]{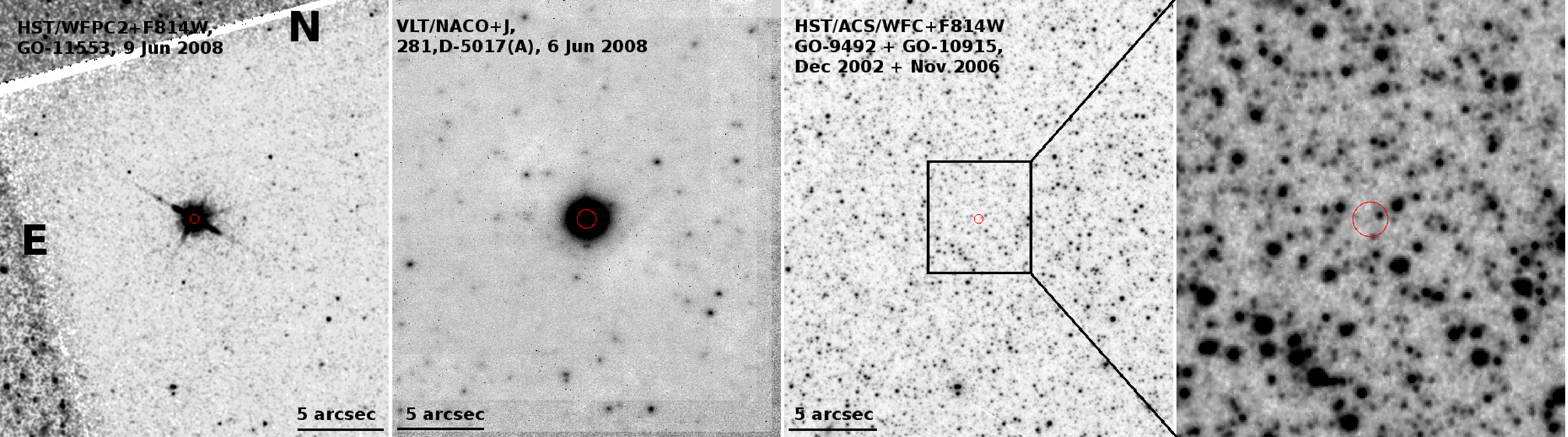}
\end{center}
\figcaption{
Four images showing the site of the NGC~300 OT\null. The first two show
the OT in outburst in 2008 June and were taken with \HST/WFPC2 and the VLT/NaCo
AO camera, using $I$ (F814W) and $J$ filters, respectively. The third frame
is a stacked \HST\/ F814W image created from pre-outburst ACS/WFC images taken
in 2002 and 2006. Small red circles show the location of the OT\null. The fourth
frame is a magnified view of the pre-outburst image, showing the location of the
transient, which we have determined to $\pm$$0\farcs01$. There is no source
detected at the OT site to a limiting F814W magnitude of 26.6. North is at the
top and east is on the left. Scale bars showing $5''$ are shown in the first
three images, and the fourth image is $6''\times7''$. The red circle in the
fourth frame has a radius of $0\farcs25$ (2.3~pc at the distance of NGC~300).
}
\end{figure}


\begin{references}

\frenchspacing

\reference{} Af{\c s}ar, M., \& Bond, H.~E.\ 2007, \aj, 133, 387 

% \reference{} Anderson, J., \& King, I.\ R. 2000, \pasp, 112, 1360

\reference{} Berger, E., \& Soderberg, A.\ 2008, The Astronomer's Telegram,
1544, 1 

\reference{} Bond, H. E. 2007, in The Nature of V838~Mon and its Light
Echo, ed.\ R. L. M. Corradi \& U. Munari (San Francisco: ASP), 130

\reference{} Bond, H. E., et al.\ 2003, \nat, 422, 405

\reference{} Bond, H.~E., \& Siegel, M.~H.\ 2006, \aj, 131, 984 

\reference{} Bond, H.~E., Walter, F.~M., \& Vel\'asquez, J.\ 2008, \iaucirc,
8946, 2 

\reference{} Corradi, R. L. M., \& Munari, U., eds., 2007, The Nature of V838
Monocerotis and its Light Echo (San Francisco: ASP)

\reference{} Davies, B., Oudmaijer,  R.~D., \& Sahu, K.~C.\ 2007, \apj, 671,
2059 

% \reference{} de Jager, C.\ 1998, \aapr, 8, 145 

\reference{} DePoy, D.~L., et al.\  2003, \procspie, 4841, 827 

% \reference{} Deguchi, S., Nakashima,  J.-i., Kwok, S., \& Koning, N.\ 2007,
% \apj, 664, 1130 

\reference{} Gieren, W., Pietrzy{\'n}ski, G., Soszy{\'n}ski, I., Bresolin, F.,
Kudritzki, R.-P.,  Minniti, D., \& Storm, J.\ 2005, \apj, 628, 695 

\reference{} Humphreys, R.~M., \& Davidson, K. 1994, \pasp, 106, 1025

\reference{} Humphreys, R.~M.,  Davidson, K., \& Smith, N.\ 1999, \pasp, 111,
1124 

\reference{} Humphreys, R. M., Davidson, K., \& Smith, N. 2002, \aj, 124, 1026

\reference{} Humphreys, R.~M., et al.\ 2006, \aj, 131, 2105 

\reference{} Jones, T.~J., et al.\  1993, \apj, 411, 323 

\reference{} Kulkarni, S.~R., et al.\ 2007, \nat, 447, 458

\reference{} Landolt, A. U. 1992, \aj, 104, 340

\reference{} Lenzen, R., et al. 2003, \procspie, 4841, 944 

% \reference{} Li, W., Filippenko, A.~V.,  Van Dyk, S.~D., Hu, J., Qiu, Y.,
% Modjaz, M.,  \& Leonard, D.~C.\ 2002, \pasp, 114, 403 

\reference{} Maeder, A., \& Meynet, G.\ 2000, \araa, 38, 143 

% \reference{} Marcelin, M., Boulesteix, J., \& Georgelin, Y. P. 1985, \aap, 151,
% 144

\reference{} Marshall, J.~L., et  al.\ 2008, \procspie, 7014,  

\reference{} Monard, L.~A.~G.\ 2008, \iaucirc, 8946, 1 

% \reference{} Pastorello, A., et  al.\ 2007, \nat, 449, 1 

\reference{} Patel, M., Oudmaijer,  R.~D., Vink, J.~S., Bjorkman, J.~E.,
Davies, B., Groenewegen, M.~A.~T.,  Miroshnichenko, A.~S., \& Mottram, J.~C.\
2008, \mnras, 385, 967 
  
\reference{} Prieto, J.~L.\ 2008, The  Astronomer's Telegram, 1550, 1 

\reference{} Prieto, J.~L., et al.\  2008, \apjl, 681, L9 

\reference{} Rau, A., Kulkarni, S.~R., Ofek, E.~O., \& Yan, L.\ 2007, \apj, 659,
1536 

\reference{} Rizzi, L., Bresolin, F., Kudritzki, R.-P., Gieren, W., \&
Pietrzy{\'n}ski, G.\ 2006, \apj, 638, 766 

\reference{} Rousset, G., et al. 2003, \procspie, 4839, 140

\reference{} Smartt, S.~J., Eldridge,  J.~J., Crockett, R.~M., \& Maund, J.~R.\
2008, arXiv:0809.0403 

\reference{} Smith, N., Ganeshalingam,  M., Li, W., Chornock, R., Steele, T.~N.,
Silverman, J.~M., Filippenko,  A.~V., \& Mobberley, M.~P.\ 2008,
arXiv:0811.3929 

\reference{} Sparks, W.~B., et al.\  2008, \aj, 135, 605 

\reference{} Steele, T.~N.,  Silverman, J.~M., Ganeshalingam, M., Lee, N., Li,
W.,  \& Filippenko, A.~V.\ 2008, Central Bureau Electronic Telegrams, 1275, 1 

\reference{} Thompson, T.~A.,  Prieto, J.~L., Stanek, K.~Z., Kistler, M.~D.,
Beacom, J.~F.,  \& Kochanek, C.~S.\ 2008, arXiv:0809.0510 

\reference{} Tylenda, R., \& Soker, N.\ 2006, \aap, 451, 223 

\reference{} Van Dyk, S.~D.\ 2005, The Fate of the Most Massive Stars, 332, 47 

\reference{} Van Dyk, S.~D., Peng,  C.~Y., King, J.~Y., Filippenko, A.~V.,
Treffers, R.~R., Li, W.,  \& Richmond, M.~W.\ 2000, \pasp, 112, 1532 

\reference{} Wisniewski, J.~P., Morrison, N.~D., Bjorkman, K.~S.,
Miroshnichenko, A.~S., Gault, A.~C.,  Hoffman, J.~L., Meade, M.~R., \& Nett,
J.~M.\ 2003, \apj, 588, 486 


\end{references}
\end{document}